\documentclass[sigconf,9pt]{acmart}
\usepackage{booktabs}
\usepackage{multirow}
\usepackage{tikz}
\usepackage{amsmath}
\usepackage{siunitx}
\usepackage[abbreviations]{foreign}
\usepackage{xspace}
\usepackage[inline]{enumitem}
\usepackage{csvsimple}
\usepackage[algo2e,ruled,lined,boxed,commentsnumbered, noend, linesnumbered, procnumbered]{algorithm2e}
\usepackage{subcaption}
\usepackage{xurl}

\usepackage{pifont}%

\usepackage{listings}
\usepackage{xcolor}

\definecolor{codegreen}{rgb}{0,0.6,0}
\definecolor{codegray}{rgb}{0.5,0.5,0.5}
\definecolor{codepurple}{rgb}{0.58,0,0.82}
\definecolor{backcolour}{rgb}{0.95,0.95,0.92}

\lstdefinestyle{mystyle}{
    language=Python,
    backgroundcolor=\color{backcolour},   
    commentstyle=\color{codegreen},
    keywordstyle=\color{magenta},
    numberstyle=\tiny\color{codegray},
    stringstyle=\color{codepurple},
    basicstyle=\ttfamily\footnotesize,
    breakatwhitespace=false,         
    breaklines=true,                 
    captionpos=b,                    
    keepspaces=true,                 
    numbers=left,                    
    numbersep=5pt,                  
    showspaces=false,                
    showstringspaces=false,
    showtabs=false,                  
    tabsize=2,
    morekeywords={self},
}

\usepackage{ifthen}
\newboolean{showcomments}
\setboolean{showcomments}{false}
\ifthenelse{\boolean{showcomments}}
{ \newcommand{\mynote}[3]{
		\fbox{\bfseries\sffamily\scriptsize#1}
		{\small$\blacktriangleright$\textsf{\emph{\color{#3}{#2}}}$\blacktriangleleft$}}
	\newcommand{\zzz}[1]{{\setlength{\fboxsep}{2pt}\fcolorbox{black}{yellow}{\textsf{\emph{#1}}}}\xspace}}
{ \newcommand{\mynote}[3]{}
	\newcommand{\zzz}[1]{}}

\usepackage{acronym}
\acrodef{DL}{decentralized learning}
\acrodef{ML}{machine learning}
\acrodef{D-PSGD}{decentralized parallel stochastic gradient descent}
\acrodef{FL}{federated learning}
\acrodef{SGD}{stochastic gradient descent}
\acrodef{IID}{independent and identically distributed}
\acrodef{non-IID}{non independent and identically distributed}
\acrodef{RMSE}{root mean square error}
\acrodef{RMW}{random model walk}
\acrodef{GL}{gossip learning}
\acrodef{EL}{epidemic learning}
\acrodef{DWT}{discrete wavelet transform}
\acrodef{FFT}{fast Fourier transform}
\acrodef{MI}{mutual information}
\acrodef{DP}{differential privacy}
\acrodef{VN}{virtual node}
\acrodef{RN}{real node}
\acrodef{LDP}{local differential privacy}
\acrodef{PNDP}{pairwise network differential privacy}
\acrodef{PNLDP}{pairwise network local differential privacy}
\acrodef{GI}{gradient inversion}
\acrodef{CML}{collaborative machine learning}
\acrodef{TPR}{true positive rate}
\acrodef{FPR}{false positive rate}
\acrodef{MoE}{Mixture-of-Experts}
\acrodef{FFN}{Feed-Forward Network}
\acrodef{EP}{expert parallelism}
\acrodef{LLM}{large language model}
\acrodef{TTFT}{time-to-first-token}

\newcommand{\sys}{\textsc{HarMoEny}\xspace}

\newcommand{\deepspeed}{\textsc{DeepSpeed}\xspace}
\newcommand{\deepspeedmii}{\textsc{DeepSpeed-MII}\xspace}
\newcommand{\deepspeedmoe}{\textsc{DeepSpeed-MoE}\xspace}
\newcommand{\fastmoe}{\textsc{FastMoE}\xspace}
\newcommand{\fastermoe}{\textsc{FasterMoE}\xspace}
\newcommand{\exflow}{\textsc{ExFlow}\xspace}
\newcommand{\pytorch}{\textsc{PyTorch}\xspace}
\newcommand{\switch}{\textsc{Switch128}\xspace}
\newcommand{\qwen}{\textsc{Qwen}\xspace}
\newcommand{\bookcorpus}{\textsc{bookcorpus}\xspace}
\newcommand{\wikitext}{\textsc{wikitext}\xspace}
\newcommand{\wmt}{\textsc{wmt19}\xspace}

\graphicspath{ {figures/} }
\usepackage{tikz}
\usepackage[eulergreek]{sansmath}
\usepackage{pgfplots}
\usepackage{pgfplotstable}
\usepackage{cleveref}
\usepackage{comment}
\crefname{assumption}{assumption}{assumptions}

\pgfplotsset{compat=newest}
\usepgfplotslibrary{external,units,colorbrewer,groupplots,fillbetween,statistics,colormaps}
\tikzsetexternalprefix{figures/}
\tikzset{external/mode=list and make}
\usetikzlibrary{patterns,shapes.misc}

\makeatletter
\begingroup\endlinechar=-1\relax
\everyeof{\noexpand}%
\edef\x{\endgroup\def\noexpand\homepath{%
		\@@input|"kpsewhich --var-value=HOME" }}\x
\makeatother

\def\overleafhome{/tmp}
\newcommand{\inputplot}[2]{%
	\ifx\homepath\overleafhome%
	\IfBeginWith{#1}{plots}{\includegraphics{main-figure#2.pdf}}{#1}%
	\else%
	{\sffamily\scriptsize\input{#1}}
	\fi
}

\newcommand{\newgroupwidth}[2]%
{\expandafter\xdef\csname groupwidth#1\endcsname{#2}}

\newcounter{groupwidth}
\newsavebox{\groupwidthbox}
\makeatletter
{\edef\groupnumber{#1}%
	\stepcounter{groupwidth}%
	\@ifundefined{groupwidth\thegroupwidth}{\pgfmathsetlengthmacro{\mywidth}{\linewidth/\groupnumber}}%
	{\expandafter\let\expandafter\mywidth\csname groupwidth\thegroupwidth\endcsname}%
	\begin{lrbox}{\groupwidthbox}%
		\tikzset{/pgfplots/width={\mywidth}}%
		\ignorespaces}%
	{\end{lrbox}%
	\usebox\groupwidthbox
	\pgfmathsetlengthmacro{\mywidth}{\mywidth + (\linewidth - \wd\groupwidthbox)/\groupnumber}
	\immediate\write\@auxout{\string\newgroupwidth{\thegroupwidth}{\mywidth}}}
\makeatother

\usepackage{amsmath,amssymb,amsfonts,amsthm}
\usepackage{physics}
\usepackage{enumitem}
\usepackage{thm-restate}
\usepackage{bbm}
\theoremstyle{definition}

\theoremstyle{remark}

\allowdisplaybreaks

\usepackage{algorithm}
\usepackage{algpseudocode}

\copyrightyear{2025}
\acmYear{2025}

\setcopyright{none}

\ccsdesc[500]{Computer systems organization~Parallel architectures}
\ccsdesc[500]{Computer systems organization~Distributed architectures}

\makeatletter
\AtBeginDocument{%
	\def\ltx@label#1{\cref@label{#1}}%
	
	\def\label@in@display@noarg#1{\cref@old@label@in@display{#1}}%
	
	\def\label@in@mmeasure@noarg#1{%
		\begingroup
		\measuring@false
		\cref@old@label@in@display{#1}%
		\endgroup
	}%
}
\makeatother

\begin{document}

\title{\sys: Efficient Multi-GPU Inference of MoE Models}
\author{Zachary Doucet}
\affiliation{
  \institution{McGill}
  \country{Canada}
}

\author{Rishi Sharma}
\affiliation{
  \institution{EPFL}
  \country{Switzerland}
}

\author{Martijn de Vos}
\affiliation{
  \institution{EPFL}
  \country{Switzerland}
}

\author{Rafael Pires}
\affiliation{
  \institution{EPFL}
  \country{Switzerland}
}

\author{Anne-Marie Kermarrec}
\affiliation{
  \institution{EPFL}
  \country{Switzerland}
}

\author{Oana Balmau}
\affiliation{
  \institution{McGill}
  \country{Canada}
}

\begin{abstract}

Mixture-of-Experts (MoE) models offer computational efficiency during inference by activating only a subset of specialized experts for a given input.
This enables efficient model scaling on multi-GPU systems that use expert parallelism without compromising performance.
However, load imbalance among experts and GPUs introduces waiting times, which can significantly increase inference latency.
To address this challenge, we propose \sys, a novel solution to address MoE load imbalance through two simple techniques: \emph{(i)} dynamic token redistribution to underutilized GPUs and \emph{(ii)} asynchronous prefetching of experts from the system to GPU memory.
These techniques achieve a near-perfect load balance among experts and GPUs and mitigate delays caused by overloaded GPUs.
We implement \sys and compare its latency and throughput with four MoE baselines using real-world and synthetic datasets.
Under heavy load imbalance, \sys increases throughput by 37\%--70\% and reduces time-to-first-token by 34\%--41\%, compared to the next-best baseline.
Moreover, our ablation study demonstrates that \sys's scheduling policy reduces the GPU idling time by up to 84\% compared to the baseline policies.

\end{abstract}

\keywords{ML Inference, Mixture-of-Experts Models, Load Balancing}

\maketitle

\section{Introduction}
\label{sec:intro}

Scaling \ac{ML} models to billions of parameters has enabled powerful generative models~\cite{kaplan2020scalinglawsneurallanguage, yang2024harnessing}.
One of the main applications of these models is natural language processing, where \acfp{LLM} such as GPT-3~\cite{brown2020language} and GPT-4~\cite{achiam2023gpt} are widely used for tasks like text generation and question answering.
However, these applications come with steep costs and high energy consumption.
As model sizes grow, inference becomes increasingly expensive \cite{bianchini2024datacenter}, and it already constitutes the majority of ML workloads. 
NVIDIA and AWS estimate that up to 90\% of the ML workloads are serving deep neural network models~\cite{nvidia-inference, aws-inference}.
To address this critical challenge, this paper focuses on strategies for efficient \ac{ML} inference.

One promising approach is the use of \Acf{MoE} models~\cite{jacobs1991adaptive}.
Compared to traditional models, \ac{MoE} models can provide more than 10$\times$ reduction in computation requirements for inference, without sacrificing accuracy~\cite{lepikhin2021gshard,fedus2022switchtransformers}.
The Switch Transformers~\cite{fedus2022switchtransformers}, Mixtral~\cite{jiang2024mixtralexperts}, Qwen~\cite{yang2024qwen2}, and DeepSeek~\cite{liu2024deepseek} model families are some of the most successful \acp{MoE}.
Each expert in an \ac{MoE} model is trained to focus on a specific subset of tasks or data patterns.
A gating mechanism, often a smaller neural network called a \textit{router}, decides which experts will process a given input.
This approach allows \ac{MoE} models to scale, using only a fraction of their capacity per input, resulting in computational savings. %

\begin{figure}[b!]
	\centering
	\inputplot{plots/expert_imbalance}{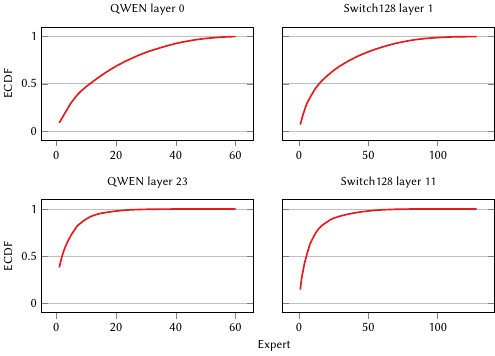}
	\caption{The ECDF of token placement across all 128 experts on the \bookcorpus dataset for the \qwen and \switch models. The workload exhibits strong skew where a handful of experts receive over half the total number of tokens. Efficiently handling pronounced, dynamic skew is a challenge for existing MoE serving systems.}
	\label{fig:expert_imbalance}
\end{figure}

Despite the benefits in terms of computation, \ac{MoE} models have a significant memory footprint.
While only a fraction of experts are activated per input, all the experts need to be available in GPU memory to process a batch of requests.
Therefore, serving \ac{MoE} models requires a combination of multiple GPUs, each holding a subset of the experts, also known as expert parallelism (EP).

Serving \ac{MoE} models using multiple GPUs has two significant bottlenecks that increase the end-to-end request latency: \emph{(i)} synchronization, and \emph{(ii)} load imbalance among the GPUs~\cite{yao2024exploiting,he2022fastermoe,he2021fastmoe,li2023accelerating}.
First, \acp{MoE} models require two synchronization steps (all-to-all communication) between GPUs in every \ac{MoE} block.
Based on the decisions of the router in each \ac{MoE} block, each GPU first scatters its inputs to the relevant experts on other GPUs and gathers them back after the computation is done. 
These synchronizations are essential to ensure that all GPUs have the complete results from the current \ac{MoE} block before moving on to the next one. Recent \ac{MoE} training and inference frameworks have alleviated this slowdown by optimizing the all-to-all communication~\cite{he2022fastermoe,li2023accelerating, yao2024exploiting}. They either resort to replication~\cite{he2022fastermoe} of popular experts or collocate popular experts across \ac{MoE} layers~\cite{li2023accelerating, yao2024exploiting}.
In these cases, the popularity of experts is determined through profiling and scheduling performed offline.
However, expert popularity is dynamic and depends on the tokens in the input request.
The tokens depend on the domain of the input request, \ie, a medical prompt activates different tokens when compared to a StackOverflow question~\cite{jiang2024mixtralexperts}.

Indeed, depending on the workload, some experts are significantly more \emph{popular} than others and are assigned more tokens than other experts.
We show this in \Cref{fig:expert_imbalance} for the \qwen and \switch \ac{MoE} models during a run with the \bookcorpus dataset.
Token placements across experts are non-uniform, and tokens are disproportionately routed to a small subset of experts per input.
For instance, in layer 0 of the \qwen model, only three experts receive an average of 19\% of the tokens, whereas in the final layer, three experts receive as much as 60\% on average.
This skew becomes more pronounced in deeper layers of the model. Moreover, the expert skew is dynamic, with the token distribution varying across queries.
This skew in expert popularity leads to a load imbalance among the GPUs and to significant GPU under-utilization.

\begin{figure}[t!]
	\centering
	\inputplot{plots/dataset_per_gpu4}{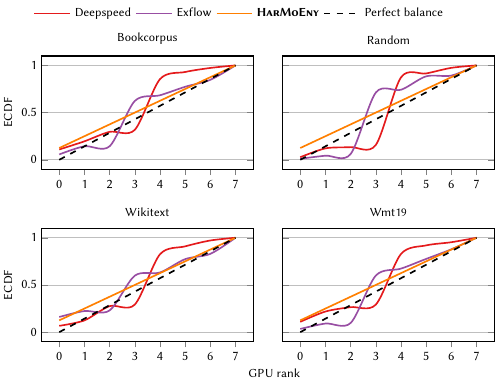}
	\caption{The ECDF of token placement across GPUs in an 8-GPU NVIDIA DGX machine, over four datasets, for a Switch Transformer model~\cite{fedus2022switchtransformers} with 128 experts. \sys achieves near-perfect load balancing.
		}
	\label{fig:dataset_per_gpu4}
    \vspace{-5mm}
\end{figure}

In this paper, we target load imbalance caused by skewed expert popularity to improve the efficiency of multi-GPU \ac{MoE} inference. Expert popularity imbalances (shown in Figure~\ref{fig:expert_imbalance}) lead to significant imbalances in GPU use. %
Figure~\ref{fig:dataset_per_gpu4} shows the load imbalance across GPUs for two popular \ac{MoE} load balancing approaches and our solution, \sys.
\deepspeed uses a round-robin distribution of tokens to GPUs (also used by others such as \fastmoe~\cite{he2021fastmoe}, and \fastermoe~\cite{he2022fastermoe}).
Another compelling approach used by \exflow~\cite{yao2024exploiting} is workload profiling and integer programming to determine the optimal expert placement.
This technique yields better load balance at times, but is not fast enough to adapt to skew changes across batches.
Both approaches yield an imbalanced distribution of tokens to experts, leading to high GPU idle times, as we will show in Section~\ref{sec:load_imbalance}.
We found that this idle time can take up to 86\% of the time for GPUs housing unpopular experts. In contrast, \sys achieves almost perfect load balance.

\sys uses two simple, yet powerful techniques to achieve near-perfect load balance among GPUs: \textit{token rebalancing} from overutilized to underutilized GPUs, and \textit{asynchronous prefetching of experts} from system to GPU memory.
\sys adapts on the fly to changes in expert popularity with no drops in throughput and does not need any profiling.

To achieve this, \sys modifies the MoE logic that is replicated on all GPUs.
During each batch, the GPUs exchange a summary of the token distribution to get a global vision of the token-to-GPU and token-to-expert assignment.
The metadata in this step is small (4kB) and introduces minimal overheads.
Then, each GPU can deterministically infer \textit{the same} token-to-expert and token-to-GPU schedule in parallel, with no further synchronization needed.
\sys balances the token load such that some of the tokens that were destined to overutilized GPUs (\ie, hosting more popular experts) are rerouted to underutilized GPUs. To ensure that the rerouted tokens can be processed by the popular experts once they are rebalanced to the underutilized GPUs, \sys uses asynchronous expert prefetching. 
\sys can look ahead and make sure that the right experts are paged into GPU memory from the system memory, and the ones that are not needed anymore are discarded.
Swapping out an expert from GPU memory only requires an overwrite.

\sys uses prefetching and load rebalancing, both well-established techniques in datacenter scheduling~\cite{delimitrou2014quasar,tirmazi2020borg}.
However, we are the first to adapt such scheduling techniques to \ac{MoE} models and show they eliminate GPU idleness almost completely.
We compare \ac{MoE} inference in \sys to four state-of-the-art \ac{MoE} systems: \exflow~\cite{yao2024exploiting}, \fastmoe~\cite{he2021fastmoe} and \fastermoe~\cite{he2022fastermoe}, and \deepspeed-\textsc{Tutel}~\cite{rajbhandari2022deepspeed}.
We show that in workloads with skewed expert popularity, \sys is up to 41.1\% faster than the next-best baseline, in terms of time-to-first-token.
Furthermore, \sys maintains stable and low inference latency even as skew and expert popularity changes.
Thanks to its lightweight techniques, \sys has the capacity to quickly adapt to various datasets, which is a drawback of profiling-based approaches.

\vspace{1mm}
\noindent\textbf{Contributions.} This paper makes the following contributions:

\begin{enumerate}
    \item We empirically study the compute utilization of a GPU cluster running \ac{MoE} inference and conclude that expert popularity imbalance has a much higher impact on inference latency than all-to-all synchronization (\Cref{sec:load_imbalance}).
    With a balanced load across GPUs, the all-to-all synchronization accounts for only 2\% of the total execution time.
    \sys's design follows from this observation. 
    \item We design and implement \sys (\Cref{sec:design}).
    \sys uses two complementary techniques to achieve almost perfect load balancing: token rebalancing and asynchronous prefetching of experts.
    \sys is open source\footnote{See \url{https://github.com/sacs-epfl/HarMoEny}.} and implemented on top of \textsc{PyTorch}.
    \item We evaluate \sys with real datasets and synthetic benchmarks, showing that \sys maintains a low and steady inference latency in fluctuating workloads, with different skew levels, and across different datasets (\Cref{sec:evaluation}).
    Our simple and efficient approach reduces the waiting time of GPUs in the all-to-all synchronization step by up to 84.7\% compared to baseline policies.
    
\end{enumerate}

\section{Background}
\label{sec:background}

\begin{figure}[t]
    \centering
    \includegraphics[width=0.9\linewidth]{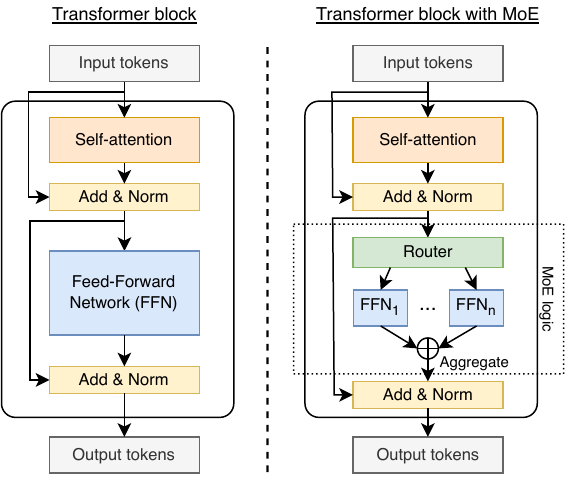}
    \caption{A transformer block (left) and a \ac{MoE} block, containing the \ac{MoE} logic (right). \ac{MoE}-based models replace the \ac{FFN} with a router and multiple experts implemented as \acp{FFN}.}
    \label{fig:transformer}
\end{figure}

\noindent\textbf{Transformer models} are nowadays widely used for \ac{ML} tasks~\cite{vaswani2017attention}.
A typical transformer model consists of multiple transformer blocks, each designed to process tokens through self-attention mechanisms and \acfp{FFN}.
A \textit{token} here refers to an intermediate value representing a single element, \eg, a word~\cite{bengio2000neuralprobabilisticlanguagemodel}, a sub-word~\cite{kudo2018subwordregularizationimprovingneural}, or a character~\cite{gupta2019characterbasednmttransformer}.
As illustrated in \Cref{fig:transformer} (left), a transformer block takes some tokens as input and consists of two main components: a self-attention mechanism and an \ac{FFN}.
The self-attention mechanism captures relationships between input elements across the sequence, allowing the model to focus on different parts of the input simultaneously.
These outputs are passed into an \ac{FFN}, which applies two dense layers with an activation function in between to refine the representations.
The \ac{FFN} is the most time-consuming part of the transformer block~\cite{xu2024resource}.
The resulting output tokens are then forwarded to the subsequent transformer block.

\noindent\textbf{Mixture-of-Experts (MoEs)} is a type of sparse computation that selectively activates only parts of the network, called \textit{experts}~\cite{jacobs1991adaptive}.
A transformer block using \ac{MoE} is shown in~\Cref{fig:transformer} (right).
In a \ac{MoE} model, some or all transformer blocks can have the \ac{MoE} layer.
We refer to a transformer block having the \ac{MoE} layer as a \emph{\ac{MoE} block}.
In contrast to a typical transformer block with a single \ac{FFN}, an \ac{MoE} layer contains multiple experts, implemented by smaller \acp{FFN}, with each expert having its own set of weights~\cite{shazeer2017outrageouslylargeneuralnetworks}.
The number of experts in each block is typically between 8--128~\cite{fedus2022switchtransformers}.
MoEs assign each token to only a portion of the network, \ie, a subgroup of experts rather than passing it through the entire model.

\begin{figure}[b]
    \centering
    \includegraphics[width=\linewidth]{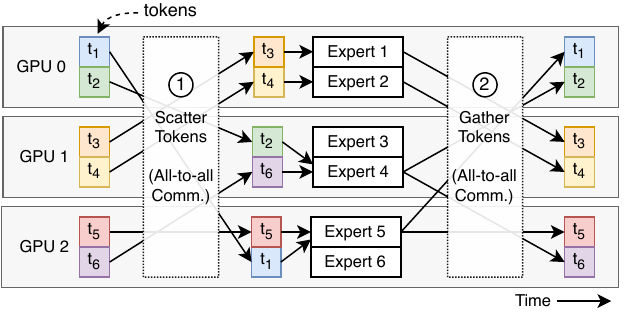}
    \caption{Token scattering (step 1) and gathering (step 2) when using expert parallelism. Experts are split across GPUs. One batch requires two all-to-all synchronization barriers.
    }
    \label{fig:moe}
\end{figure}

The assignment of tokens to experts is managed by the \textit{router}, a component responsible for directing each token to a subset of the model (in green in \Cref{fig:moe}). 
The router is usually implemented as a trainable function~\cite{shazeer2017outrageouslylargeneuralnetworks}, optimized through backpropagation to discover productive token-to-expert assignments.
To prevent bottlenecks, a loss term during training encourages the router to distribute tokens evenly across the experts.
The router assigns each expert a value between 0 and 1.
Then, expert assignment is handled with a \textit{top-$k$} strategy—directing the token to the $k$ experts with the highest values, with each expert’s output weighted accordingly. The parameter
 $k$ is usually set to 1 or 2, as higher values quickly raise costs while offering diminishing returns~\cite{fedus2022switchtransformers, jiang2024mixtralexperts}.
After processing the expert \ac{FFN}, the token’s output is combined, normalized, and passed to the next layer.

\noindent\textbf{\Acf{EP}.}
In multi-GPU scenarios, transformers typically use three types of parallelism: \emph{(i)} data parallelism, where input data is sharded across the GPUs~\cite{dean2012}, \emph{(ii)} model parallelism, where different parts of the model are split across the GPUs at a layer- or component-granularity~\cite{krizhevsky2009learning}, and \emph{(iii)} tensor parallelism, where large tensors (\ie, matrix operations within a single layer) are split across GPUs~\cite{shoeybi2020megatronlmtrainingmultibillionparameter}.
\Acf{EP}~\cite{he2021fastmoe} combines aspects of data parallelism and model parallelism to support \acp{MoE}.

With \ac{EP}, the self-attention and router layers are replicated across GPUs (data parallelism), but each GPU only loads a subset of experts, distributing the complete set of experts across all GPUs~\cite{fedus2022switchtransformers} (model parallelism).
During the forward pass, each GPU receives a minibatch of the input request comprising a set of input tokens.
All the GPUs independently compute self-attention on their minibatches in parallel.
Here, the routers in each GPU assign the tokens from the minibatches to experts.
Since the tokens and their assigned experts can potentially be on different GPUs, an all-to-all scatter communication step ensures that each GPU receives the tokens destined for the experts it hosts.
This step introduces the first synchronization barrier in \ac{MoE} blocks.
Upon receiving the correct tokens, each GPU performs the expert computation on the received tokens using the experts it holds.
After the expert computation, an all-to-all gather communication step returns the computed results for each token back to the GPUs responsible for their corresponding inputs.
Finally, the resulting tokens are then used as input for the next \ac{MoE} block.

Figure~\ref{fig:moe} shows an example of an \ac{EP} assignment with 6 experts split across 3 GPUs demonstrating the expert computations and the all-to-all communications.
Experts 1 and 2 are housed in GPU 0, experts 3 and 4 to GPU 1, and experts 5 and 6 to GPU 2.
The self-attention mechanism outputs tokens $t_1$ to $t_6$.
When GPU 0 processes its batch, the router assigns tokens to experts—sending $a$ tokens to expert 1, $b$ to expert 2, $c$ to expert 3, and so on.
Since GPU 0 only houses experts 1 and 2, it must send tokens for experts 3 and 4 to GPU 1 and tokens for experts 5 and 6 to GPU 2 (step 1).
Once GPU 0 completes computations for its experts, it returns the processed tokens to their originating GPUs through another all-to-all communication, posing another synchronization (step 2).

\section{Effect of load imbalance in \ac{MoE} Inference}
\label{sec:load_imbalance}

\ac{MoE} routers are trained to balance the token load across the experts~\cite{fedus2022switchtransformers,jiang2024mixtralexperts}.
However, during inference, the expert selection, and hence the computation load across GPUs is often skewed as shown in \Cref{fig:dataset_per_gpu4}.
We now show the impact of skewed expert popularity and load imbalance on performance.

\begin{figure}
	\centering
	\inputplot{plots/no_rebalancing}{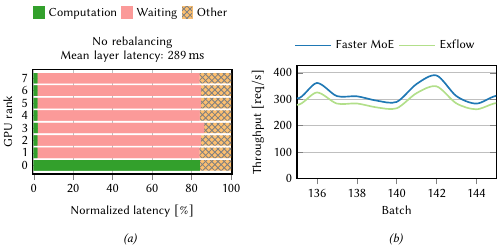}
	\caption{\emph{(a)} Fixed expert placement causes long waiting due to load imbalance. \emph{(b)} Throughput fluctuates with static placement, given the nature of requests. 
	}
	\label{fig:no_rebalancing}
\end{figure}

We first show that \emph{expert popularity skew causes significant GPU under-utilization} leading to high end-to-end latency, in line with prior work. \Cref{fig:no_rebalancing}~(a) presents the time breakdown of GPU computations when serving a Switch transformer model with 6 layers and \num{128} experts.
We inject artificial token skew such that $90\%$ of the tokens are assigned to the first 10 experts.
We run the experiment with 8x V100 GPUs  (see \Cref{sec:experimental_setup} for the full experimental setup).
In this workload, GPU 0 is assigned the most popular experts. Due to the all-to-all synchronization steps, GPUs 1--7 remain idle for more than 82\% of the time. 
In this situation, balancing the token load across GPUs and consequentially minimizing the waiting times would lead to efficient serving and reduced end-to-end latency of inference requests.

Prior work improved the efficiency of \ac{MoE} model serving in the case of static or mostly-static expert popularity~\cite{he2021fastmoe, he2022fastermoe, yao2024exploiting, li2023accelerating}.
Simple replication of the popular experts onto the under-utilized GPUs as proposed by \fastermoe~\cite{he2022fastermoe} suffices for balancing loads in workloads with static expert popularity.
In scenarios where expert popularity skew slowly changes over time, profiling can be done over time and expert placement can be adjusted periodically as done in \exflow~\cite{yao2024exploiting}.

Realistically, however, the skew and consequently the load imbalance fluctuate between batches.
This variation arises because the skew depends on the token distribution within the input requests, which is in turn influenced by the domain of the requests (\eg, medical, programming, etc.)~\cite{jiang2024mixtralexperts}.
The skew fluctuation results in unstable throughput across batches for static solutions like \fastmoe.
Furthermore, profiling-based solutions are inefficient for \ac{MoE} serving in the presence of dynamic skew due to their cost.
For instance, in our test bed, profiling and readjusting the schedule with integer programming in \exflow takes about 8.5 minutes for the Switch transformer and as much as 45 minutes for experiments with the Qwen \ac{MoE} model.
In contrast, even with low GPU utilization, the mean time to process a single batch of requests through a \ac{MoE} block is only 289 milliseconds as shown in \Cref{fig:no_rebalancing}~(a).
Therefore, \exflow does not have enough time to adapt to the dynamic expert popularity.

\Cref{fig:no_rebalancing}~(b) shows the impact on the throughput of \fastermoe~\cite{he2022fastermoe} and \exflow~\cite{yao2024exploiting} in a longer run of the same setup as \Cref{fig:no_rebalancing}~(a) with dynamically fluctuating expert-token skew across batches.
\fastermoe starts with a round-robin placement of experts on GPUs and \exflow uses 40 samples from the \bookcorpus dataset to create a schedule for expert placement on the GPUs using integer programming (see \Cref{sec:experimental_setup} for more details).
We can see that \exflow's and \fastermoe's throughput fluctuates across batches and drops by up to 37.6\% within just two consecutive batches.
Both systems have similar performance as neither system has time to adapt to the rapid fluctuations in the skew.

\begin{algorithm2e}[t]
\DontPrintSemicolon
\caption{\sys MoE Layer}
\label{alg:harmoeny}
\SetKwProg{Fn}{Procedure}{:}{end}
\SetKwInOut{Input}{Require}
\SetKwFunction{Forward}{forward}
\SetKw{KwSet}{set}
\Input{$G$: Set of GPUs.}
\;
\Fn{\textup{\textsc{forward}}($ x $)} {
    \textbf{$ // $ Step 1: token routing} \;
    $ m_{expert} \leftarrow$ \textsc{router}($x$)\label{line:token_routing} \;
    \;
    \textbf{$ // $ Step 2: metadata exchange} \;
    \textsc{sendMetadataToGPUs}($ m_{expert} $) \;
    \textbf{receive} $ m_{all}[$i$] $ from each GPU $ i \in G $ \;
    \;
    \textbf{$ // $ Step 3: token scheduling}\;
    $S_{initial} \leftarrow $ \textsc{initialAssign($ m_{all} $)}\label{line:initial_assign} \;
    $S \leftarrow $ \textsc{rebalance($ S_{initial} $)} \Comment{See \Cref{sec:scheduler}}\label{line:rebalance} \;
    \;
    \textbf{$ // $ Step 4: scatter tokens} \Comment{Step 1 in \Cref{fig:moe}} \;
    \textsc{sendTokensToGPUs}($x$, $m_{expert}$, $ S $)\\
    \textbf{receive} $ x' $ from all other GPUs\\
    \;
    \textbf{$ // $ Step 5: expert processing and async. loading} \;
    $ x'' \leftarrow $ \textsc{experts}($ S $, $ x' $) \Comment{See \Cref{sec:async_fetch}}\label{line:call_experts} \;
    \;
    \textbf{$ // $ Step 6: gather tokens} \Comment{Step 2 in \Cref{fig:moe}} \;
    \textsc{sendTokensBackToGPUs}($ S $, $x''$)\\
    \textbf{receive} $ y[i] $ from each GPU $ i  \in G $\\
    $ x \leftarrow $ \textsc{reconstruct}($ S $, $ y $, $m_{all}$)\label{line:reconstruct}\\
    
    \Return{x}\\
}
\end{algorithm2e}

\section{\sys design}
\label{sec:design}
Based on our findings in~\Cref{sec:load_imbalance}, we build \sys as a system to reduce inference latency for \ac{MoE} models.
\sys is composed of two main components:
\begin{enumerate*}[label=\emph{(\roman*)}] 
\item a scheduler that load-balances tokens across experts, 
and \item the expert pre-fetching protocol that asynchronously prefetches an expert into GPU memory.
These two techniques reduce GPU idle time introduced by token load imbalance without any online profiling and allow \sys to adapt to rapid workload fluctuations.
\end{enumerate*}
We first explain the overall workflow of \sys, and then explain each of these two components.

\subsection{\sys workflow}
\Cref{alg:harmoeny} shows the operations during a forward pass through the \ac{MoE} logic with \sys.
This \textsc{forward} function is executed by each GPU and takes some input tokens $ x $, which is a tensor of shape [batch size, sequence length, hidden dimension]. 
We assume that input tokens $ x $ have already been passed through the self-attention layer and will now be routed to and processed by the relevant experts.
This proceeds in the following six steps.

\vspace{1mm}
\noindent\textbf{Step 1: token routing.}
Input tokens $ x $ are first assigned to specific experts based on their characteristics using a router mechanism (\Cref{line:token_routing}).
This assignment creates a token-to-expert mapping, $ m_{expert} $, which is a tensor of integers representing the target expert for each token.

\vspace{1mm}
\noindent\textbf{Step 2: metadata exchange.}
Next, all GPUs exchange their computed token-to-expert distribution.
Specifically, each GPU broadcasts its local token-to-expert assignments, allowing all GPUs to build a global, shared understanding of the token distribution.
This metadata exchange step requires very little communication (a few kilobytes) and thus is efficient.
This information is stored in the array $ m_{all} $ which tracks the token-to-expert assignment for all other GPUs.
This step ensures that \sys token scheduling operates with a complete view of the workload.
We note that this metadata exchange is unique to \sys, and we experimentally show in \Cref{sec:evaluation} that its overhead is negligible.

\vspace{1mm}
\noindent\textbf{Step 3: token scheduling.}
Based on the token-to-expert assignments in $ m_{all} $, an initial (naive) token schedule $ S_{initial} $ is generated (\Cref{line:initial_assign}).
$ S_{initial} $ is a three-dimensional tensor of integers where each entry represents the number of tokens assigned from a source GPU to a destination GPU for a particular expert.
Since this might result in load imbalance issues, \sys then \emph{rebalances} this schedule through an algorithm that redistributes tokens from overburdened GPUs to underutilized ones (\Cref{line:rebalance}, also see \Cref{sec:scheduler}).

\vspace{1mm}
\noindent\textbf{Step 4: scatter tokens.}
Based on the rebalanced schedule ($ S $) and the local token-to-expert assignment $ m_{expert} $, each GPU now sends its input tokens in $ x $ to the designated GPUs through an all-to-all communication step.
Each GPU $ i $ then receives all tokens $ x' $ from other GPUs that should be processed by the receiving GPU. %

\vspace{1mm}
\noindent\textbf{Step 5: expert processing and asynchronous loading.}
The tokens in $ x' $ are then processed by the appropriate experts, resulting in $ x'' $ (\Cref{line:call_experts}).
It might be that our rebalancing algorithm assigns tokens to experts that are not currently housed by a particular GPU.
\sys employs a novel, asynchronous expert pre-fetching protocol to ensure that the required expert weights are loaded into GPU memory without delaying token processing (see \Cref{sec:async_fetch}).
Weight transfers are overlapped with computation, reducing idle GPU time.

\vspace{1mm}
\noindent\textbf{Step 6: gather tokens.}
After expert processing is completed, each GPU sends the processed tokens $ x'' $ back to their original GPUs using a second all-to-all  communication step.
This ensures that each GPU receives the results for the tokens it initially routed.
Each GPU $ i $ stores the received tokens in $ y[i] $.
Once all tokens are collected, the received tokens are restructured to ensure that the processed tokens are aligned correctly with their original order and source (\Cref{line:reconstruct}).
This completes the processing of input tokens $ x $ by the \ac{MoE} logic, yielding the final output tokens.

\begin{figure*}[t]
	\centering
	\begin{subfigure}[b]{0.45\linewidth}
		\centering
		\includegraphics[width=\linewidth]{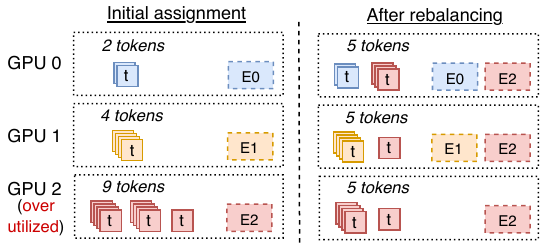}
		\caption{During rebalancing, tokens are reassigned from overutilized to underutilized GPUs.}
		\label{fig:token_scheduling_example}
	\end{subfigure}
	\hspace{0.025\linewidth}
	\begin{subfigure}[b]{0.45\linewidth}
		\centering
		\includegraphics[width=\linewidth]{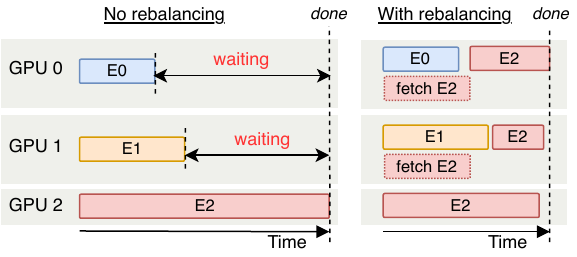}
		\caption{Timeline with operations per GPU, with and without rebalancing. Experts are fetched asynchronously (see \Cref{sec:async_fetch}).}
		\label{fig:token_scheduling_timeline}
	\end{subfigure}
	\caption{An example of the token rebalancing process by \sys. Setup:  15 input tokens, three GPUs and three experts.}
	\label{fig:token_scheduling}
\end{figure*}

\begin{algorithm2e}[t]
\DontPrintSemicolon
\caption{\sys Token Rebalancing}
\label{alg:scheduling}
\SetKwProg{Fn}{Procedure}{:}{end}
\SetKwInOut{Input}{Require}
\SetKwInOut{Output}{Output}
\Input{$G$: Set of GPUs, $E$: Set of experts, $q$: Token transfer threshold.}
\Output{Rebalanced schedule $S$}
\;
\Fn{\textup{\textsc{rebalance}}($ S_{initial} $)}{

$S \leftarrow S_{initial}$ \;
$t_{\text{avg}} \leftarrow \lfloor S.\textsc{sum}() / |G| \rfloor$ \Comment{Avg. tokens per GPU} \;
$t_g \leftarrow S.\textsc{sum}(\text{dim}=(0, 1))$ \Comment{Token count per GPU} \;

\While{\textsc{any}($t_g > t_{\text{avg}}$)} {\label{line:rebalance_loop}
    $g_{\text{max}} \leftarrow \textsc{argmax}(t_g)$\label{line:gmax} \;
    $g_{\text{from}} \leftarrow \textsc{argmax}(\textsc{sum}(S[:, :, g_{\text{max}}], \text{dim}=1))$\label{line:gfrom} \;
    $e_{\text{max}} \leftarrow \textsc{argmax}(S[g_{\text{from}}, :, g_{\text{max}}])$\label{line:emax} \;
    \;
    $t_{\text{move}} \leftarrow S[g_{\text{from}}, e_{\text{max}}, g_{\text{max}}]$\label{line:tokens_to_move} \;
    \If{$t_{\text{move}} < q$} {
        \textbf{return} $ S $ \Comment{Insufficient tokens to move} \;
    }
    \;
    $g_{\text{min}} \leftarrow \textsc{argmin}(t_g)$ \Comment{Find least loaded GPU}\label{line:find_underutilized_gpu} \;
    
    \If{$g_{\text{min}} = g_{\text{max}}$ \textbf{or} $t_g[g_{\text{min}}] + q > t_{\text{avg}}$} {
        \textbf{return} $ S $ \Comment{No feasible transfer} \;
    }
    \;
    $t_s \leftarrow \min(t_{\text{move}}, t_{\text{avg}} - t_g[g_{\text{min}}])$\label{line:actual_tokens_to_move} \;
    $S[g_{\text{from}}, e_{\text{max}}, g_{\text{max}}] \mathrel{-}= t_s$ \;
    $S[g_{\text{from}}, e_{\text{max}}, g_{\text{min}}] \mathrel{+}= t_s$ \;
    $t_g[g_{\text{max}}] \mathrel{-}= t_s$ \;
    $t_g[g_{\text{min}}] \mathrel{+}= t_s$ \;
}
\;
\textbf{return} $S$

}

\end{algorithm2e}

\subsection{Load-aware token scheduler}
\label{sec:scheduler}

One of the innovations of \sys is an efficient load-aware token scheduler that assigns each token to one of the available GPUs.
The scheduler is able to rebalance the load by re-assigning tokens that are destined to one GPU to a less crowded GPU.
We outline the algorithm in \Cref{alg:scheduling} and visualize this process in \Cref{fig:token_scheduling}.
The procedure \textsc{rebalance} takes as input an initial schedule $ S_{initial} $, which is then assigned to the variable $ S $.
$ S_{initial} $ and $ S $ are three-dimensional arrays, containing a mapping between source GPUs, experts, and destination GPUs.
Specifically, \( S[g_{\text{from}}, e, g_{\text{to}}] \) denotes the number of tokens sent from GPU $ g_{from} $ for expert $ e $ to GPU $ g_{to} $.
Thus, the first dimension corresponds to the source GPUs, the second to the experts, and the third to the destination GPUs where these experts are housed.

Imbalances arise due to uneven distribution across experts (see \Cref{sec:load_imbalance}).
We show an example in \Cref{fig:token_scheduling} which illustrates a system with 3 GPUs and \num{15} input tokens, assigned to three different experts, and using top-1 routing (\eg, in each \ac{MoE} layer, each token is processed by a single expert).
The color of each input token in \Cref{fig:token_scheduling}~(a) indicates the expert that the router has assigned to the token.
Assume that experts 0, 1, and 2 are located on GPU 0, 1, and 2, respectively.
Naively assigning input tokens to experts will result in GPU 2 having to process 9 tokens and GPU 0 just two tokens, resulting in load imbalance.
\Cref{fig:token_scheduling}~(b, left) shows a timeline with operations per GPU.
The disproportionate load on GPU 2 causes the computation time of expert 2 to grow, introducing waiting times for GPU 0 and 1 and prolonging the inference request duration.

\sys rebalances experts across GPUs by analyzing and manipulating the expert-to-GPU assignment in $ S $.
If a GPU receives more tokens than the average allocation, it is considered overutilized.
The policy then identifies the least utilized GPU and redirects as many tokens as possible to this GPU without causing it to become overutilized.
We visualize this in \Cref{fig:token_scheduling} (a, right) where expert 0 is replicated to GPU 1 as well, resulting in a situation where each GPU now processes an equal amount of tokens.
This process repeats until either all GPUs have a balanced token load, or there are no further offloading options available.

We next provide a detailed description of our greedy scheduling algorithm in \Cref{alg:scheduling}.
To this end, we first compute the average number of tokens $ t_{avg} $ and the total number of tokens $ t_g $ that each GPU has to process ($ S.$\textsc{sum}() returns the total number of tokens across all GPUs and experts).
The rebalancing loop operates iteratively to reduce load imbalances across GPUs (\Cref{line:rebalance_loop}).
This loop runs as long as there is a GPU that receives more than the average number of tokens $ t_{avg} $.
In each batch, \sys identifies the index of the most overloaded GPU, referred to as $ g_{max} $, which has the highest total token count.
The scheduler then determines the source GPU $ g_{from} $ that contributes the largest share of tokens to $ g_{max} $ (\Cref{line:gfrom}).
The term \textsc{sum}(S[:, :, $g_{max}$], \text{dim}=1) calculates the total number of tokens contributed by each source GPU to the overloaded GPU $ g_{max} $, summing over all experts.
We then identify within $ g_{from} $ the expert $ e_{max} $ that is responsible for sending the most tokens to $ g_{max} $ (\Cref{line:emax}).

Once the relevant source GPU and expert are determined, the algorithm calculates the number of tokens \( t_{\text{move}} \) to potentially transfer (\Cref{line:tokens_to_move}).
If \( t_{\text{move}} \) is smaller than some token threshold $ q $, the algorithm stops the process, as moving such a small number of tokens would not sufficiently reduce the imbalance.
The token threshold $ q $ is an important hyperparameter of \sys that decides the lower bound number of tokens necessary to offload tokens to another GPU.
The reason for introducing this threshold is to account for the time of loading an expert from memory to overlap the token processing with the expert fetching (see \Cref{fig:token_scheduling}~(b, right)).
Moving a very small number of tokens might not sufficiently reduce the imbalance to justify the cost of communication overhead, such as reconfiguring schedules and initiating transfers between GPUs.
This would result in minimal performance gains or even a net loss in efficiency. $ q $ depends on the system specifications and is independent of dynamic aspects of the workload such as expert popularity in a given batch.
We discuss how to determine this hyperparameter in \Cref{sec:async_fetch}.

If there are sufficient tokens to move, the algorithm then identifies GPU $ g_{min} $ that is assigned the least number of tokens (\Cref{line:find_underutilized_gpu}).
Tokens from $ g_{from} $ and $ e_{max} $, destined for $ g_{max} $, are then redirected to $ g_{min} $, ensuring that $ g_{min} $ does not exceed the average load $ t_{avg} $.
Specifically, the exact number of tokens transferred, $t_s$, is the smaller of \( t_{\text{move}} \) or the remaining capacity of $ g_{min} $ (\Cref{line:actual_tokens_to_move}).
After transferring tokens, the corresponding entries in \( S \) are updated to reflect the new distribution of tokens.
The total token counts in $ t_g $ are also adjusted accordingly.
This rebalancing step is repeated until either all GPUs have token counts close to $t_{avg}$ or no further feasible transfers can be made (\eg, $ t_{move} < q $).

The ECDFs in \Cref{fig:dataset_per_gpu4} illustrate the distribution of tokens assigned to each GPU across four different datasets.
\deepspeed, lacking any token load rebalancing mechanism, results in substantial load imbalances, with certain GPUs processing significantly more tokens than others.
In contrast, \sys employs an effective rebalancing algorithm that dynamically redistributes tokens, ensuring a near-uniform workload across all GPUs.
This improvement is consistent across all evaluated datasets, showcasing \sys's robustness and adaptability to different input distributions.
By mitigating load skew, \sys significantly enhances inference efficiency as we will show in \Cref{sec:evaluation}.

\subsection{Asynchronous expert fetching}
\label{sec:async_fetch}
To handle load balancing effectively, the \sys scheduler may assign tokens for certain experts to GPUs that currently do not have these experts loaded in GPU memory.
For example, \Cref{fig:token_scheduling}~(a) shows that after rebalancing, GPUs 0 and 1 get assigned some tokens designated for expert 2, which is currently not in their memory.
Thus, \sys needs a method to transfer experts into GPU memory as needed.

Simply loading experts after each expert completes processing introduces delays in the inference pipeline.
This approach is inefficient, as the GPU must wait for the current expert to be offloaded from GPU memory before loading the next expert's weights from the system memory.
Given the typically large size of expert weights (\eg, each expert in the Switch transformer and Qwen is \SI{18}{\mega\byte} and \SI{33}{\mega\byte}, respectively), this results in frequent stalls where the GPU is idle, waiting for data transfers to complete.

To achieve this efficiently, \sys prefetches experts \emph{asynchronously}, enabling transfers of experts' weights off the critical path.
Specifically, once an expert completes processing its allocated tokens, it checks for any remaining experts that need to run but are not currently loaded.
If one is found, \sys fetches the weights for this next expert from system memory, directly \emph{overwriting} the memory location of the expert that completed its processing.
We show an example of this in \Cref{fig:token_scheduling} (b, right), where GPU 0 and GPU 1 will asynchronously fetch expert 2 while computing with expert 0 and expert 1, respectively.
This technique significantly speeds up operations compared to the traditional approach of first writing the current expert to system memory and then loading the new expert into GPU memory, as the offloading to system memory is not needed. 
Our measurements show that overwriting can speedup expert loading by \textbf{5.5x}: reducing \SI{11}{\milli\second} to \SI{2}{\milli\second} for V100 GPUs.
We note that \sys benefits from asynchronous expert fetching if at least two experts fit in the GPU memory, a requirement for any system serving \acp{MoE} with many experts.

\sys's prefetching protocol relies on the preceding experts to process enough tokens so that the asynchronous weight transfer can be completed before the next expert starts.
If the computation of a particular expert is much quicker than the expert transfer time, the gains of this approach diminish.
This is influenced by the token threshold $ q $.
Thus, if $ q $ is chosen appropriately, the transfer time is effectively masked, minimizing idle periods and maintaining efficiency.

\subsection{Determining the token threshold $ q $}
\label{sec:threshold}
The token threshold $ q $ (see \Cref{sec:scheduler}) influences the number of additional experts each GPU has to load.
A small value of $ q $ causes tokens to be offloaded to GPUs without the corresponding expert and not enough tokens to amortize the cost of fetching the expert from memory.
However, a high value of $ q $ might not sufficiently address the token load imbalance and not balance the processing times of GPUs.

Ideally, we fix $ q $ such that the time to execute an expert exceeds the time to load a new expert.
Let $ |O| $ be the number of required floating operations to execute a particular expert, $\phi$ the FLOPS of the GPU being used, $ |E| $ the size of an expert in bytes, and $\beta$ the PCIe bandwidth in bytes per second.

\begin{equation}
    \label{eq:determine_q}
    \frac{|O|}{\phi} > \frac{|E|}{\beta}
\end{equation}

Experts are typically two-layer MLPs, with the first one, $ W^1$, being of size $m \times p$, and the second one, $ W^2$, being $p \times m$.
Also, let $ d_{type} $ be the size of an element in $ W^1 $ or $ W^2 $ and $ q $ the number of tokens being processed by the expert.
The expert computation can be expressed as $xW^1W^2$ where $ x $ are the input tokens.
Thus, the size $ |E| $ of an expert is given by:

\begin{equation}
    \label{eq:expert_size}
    |E| = (mp+pm)d_{type}
\end{equation}

The number of operations required to complete the expert computation with $ q $ tokens is given by:

\begin{equation}
    \label{eq:expert_computes}
    |O| = qp(2m-1)+qm(2p-1)
\end{equation}

By plugging Equations (\ref{eq:expert_size}) and (\ref{eq:expert_computes}) into \Cref{eq:determine_q}, rearranging terms, and ignoring negligible factors, we get the following inequality:

\begin{equation}
    \label{eq:final_q}
    q > \frac{\phi \cdot d_{type}}{2\beta}
\end{equation}

Since $\phi$, $d_{type}$ and $\beta$ can obtained with relative ease, \Cref{eq:final_q} guides system designers to obtain an estimate on $ q $.
A full derivation of \Cref{eq:final_q} is provided in \Cref{app:q_inequality}.
It is important to note that $q$ only depends on the system specification and the bit-precision of the parameters being served and does not depend on any dynamic properties.
Furthermore, in our experiments, we found that \sys is not extremely sensitive to $q$.
Therefore, the lower bound of \Cref{eq:final_q} provides a reliable approximation.

\section{Evaluation}
\label{sec:evaluation}

We implement \sys and compare its latency and throughput with baseline systems.

\subsection{Experimental setup}
\label{sec:experimental_setup}

We implement \sys in \num{1115} lines of code in \pytorch~\cite{torchSoftware}.
Our implementation achieves asynchronous expert fetching through a dedicated NVIDIA CUDA stream for expert loading.
The \ac{MoE} layer is written as an \texttt{nn.Module} in \pytorch for expert parallelism.
The implementation is modular and can be applied to any \pytorch model, as we demonstrate by evaluating the performance of \sys with different models.

\vspace{1mm}
\noindent\textbf{\ac{MoE} models.}
We evaluate \sys using two \ac{MoE} models: Switch Transformer~\cite{fedus2022switchtransformers} and Qwen~\cite{yang2024qwen2}.
The Switch Transformer is a language model that extends the T5 architecture~\cite{raffel2020exploring} by replacing its feed-forward layers with \ac{MoE} logic.
It contains 12 total transformer blocks alternating between a \ac{MoE} block and a classic transformer block.
Each \ac{MoE} block in this model has 128 experts and we refer to this model as \switch.
Qwen is a series of Transformer-based language models developed by Alibaba Cloud, designed to handle a wide range of tasks.
We use the Qwen 1.5 \ac{MoE} model which features \num{24} transformer blocks each of which has \num{60} experts.
We refer to this model as \qwen.
We summarize the specifications of the \ac{MoE} models used in our evaluation in \Cref{app:model_specs}.

\vspace{1mm}
\noindent\textbf{Hardware.}
All experiments are performed on a DGX1 machine featuring eight \textsc{NVIDIA} V100 GPUs (each with \SI{32}{\giga\byte} GPU memory) interconnected with \textsc{NVLink}, and \SI{500}{\giga\byte} of system memory.

\vspace{1mm}    
\noindent\textbf{Metrics.}
In our evaluation, we use two key metrics to quantify system performance: throughput and mean \acf{TTFT}.
Throughput is calculated as the total number of tokens generated across the experiment divided by the experiment length.
Mean \ac{TTFT} is a commonly-used metric that captures the average latency between the initiation of a request and the generation of the first token, reflecting the responsiveness of the system during inference.

\subsubsection{MoE System Baselines}
We compare \sys against four baselines: \deepspeed, \fastmoe, \fastermoe, and \exflow.

\textbf{\deepspeed} is a framework for distributed training and inference of large \ac{ML} models~\cite{rajbhandari2022deepspeed}.
For our evaluation, we specifically compare against \deepspeedmoe with \textsc{Tutel} enabled, which is an extension of \deepspeed and adds support for \ac{MoE} training and inference.
Furthermore, we use a high capacity factor to prevent \deepspeed from dropping tokens for a fair comparison.
Finally, we enable expert parallelism (EP) in \deepspeed which uses a round-robin placement of experts on the available GPUs.\footnote{\deepspeedmii, separate to \deepspeed, is not evaluated as the framework cannot execute on pre-Ampere GPUs, such as the V100s used in these experiments.}

\textbf{\fastmoe} is one of the earliest systems for distributed training of \ac{MoE} models~\cite{he2021fastmoe}.
The system enables large-scale \ac{MoE} training by allowing expert modules to be placed across multiple GPUs and nodes.
\fastmoe provides flexibility by allowing developers to use custom gate and expert networks, with built-in support for Transformer-based models like Megatron-LM.
It relies on optimized CUDA kernels for rapid data movement and expert selection resulting in high performance.
\fastmoe also features parallel expert computation to maximize hardware usage.

\textbf{\fastermoe} addresses token load imbalance in \fastmoe through dynamic shadowing and fine-grained scheduling, while introducing congestion-avoiding expert selection during training~\cite{he2022fastermoe}.
The system introduces dynamic shadowing, which replicates parameters of heavily used experts (popular experts) across workers, reducing the communication overhead associated with imbalanced workloads.
\fastermoe also introduces a topology-aware gating function that directs inputs to the experts with lower latency.
This function reduces communication overhead by prioritizing local expert assignments and avoiding congested network links.

\textbf{\exflow} addresses the inefficiencies of \ac{MoE} model inference by exploiting inter-layer expert affinity~\cite{yao2024exploiting}.
The key innovation lies in leveraging the observed tendency for tokens to follow predictable routing patterns across consecutive \ac{MoE} layers.
\exflow optimizes expert placement based on this affinity seen during training or a held-out dataset using an integer programming approach.
Efficient placement of experts in \exflow reduces the need for cross-GPU token routing, optimizing the duration of all-to-all communications.

\subsubsection{Datasets}
\label{sec:datasets}
Our experiments involve three real-world datasets.
\begin{enumerate*}[label=(\arabic*)]
    \item \bookcorpus is a large-scale dataset comprising up to \num{7185} unique books originally collected from smashwords.com~\cite{zhu2015moviebook}. The collection comprises multiple genres and literary styles;
    \item \wikitext is a collection of over 100 million tokens scraped from \textit{Good} and \textit{Featured} articles on Wikipedia~\cite{merity2016pointer};
    \item \wmt includes translation pairs between two languages \cite{wikimedia2019wmt19}. We work with the German-to-English set with 34.8 million translation pairs.
\end{enumerate*}

To better understand the performance of \sys and baselines, we also adopt two synthetic datasets.
\begin{enumerate*}[label=(\arabic*)]
    \item \textsc{Random} is a dataset that is constructed by stringing random tokens in a sequence of a desired length. A seed is set to ensure that random produces an identical dataset across separate runs.
    \item \textsc{Constant} is a dataset where a single token is repeated to a desired length for all batches.
\end{enumerate*}

\textbf{Expert popularity skew.}
In addition to real-world datasets, we also experiment with artificial expert popularity skews. 
To introduce artificial skew, we modify the router in each \ac{MoE} block.
We implement a configurable router skew mechanism according to the desired skew $ \alpha $, where $0 \leq \alpha \leq 1$, and the number of  experts $E$.
The selected skewed experts are assigned a probability proportional to $ \alpha $ and the remaining experts share the remaining probability evenly, ensuring that the sum of the distribution equals 1.
During routing, the router uses the multinomial distribution to sample tokens based on these probabilities, ensuring that the token distribution aligns with the desired skew.
This allows us to have fine-grained control over the load imbalance and study the performance of the baselines and \sys in more detail.

\subsection{Comparison to state-of-the-art systems}

\begin{figure}[t!]
	\centering
	\inputplot{plots/skew}{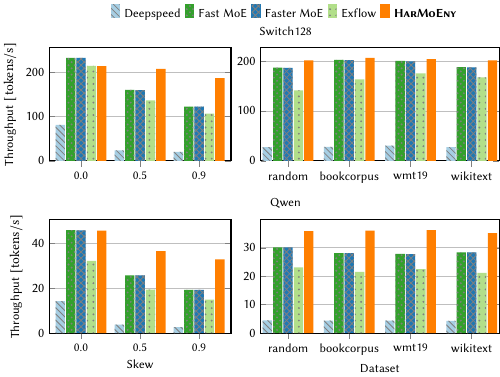}
	\caption{Throughput ($\uparrow$ is better) for different systems and different skews (left) and datasets (right) when using the \switch (top) and \qwen models (bottom).}
	\label{fig:skew-tput}
\end{figure}

\begin{figure}[t!]
	\centering
	\inputplot{plots/skew-mttft}{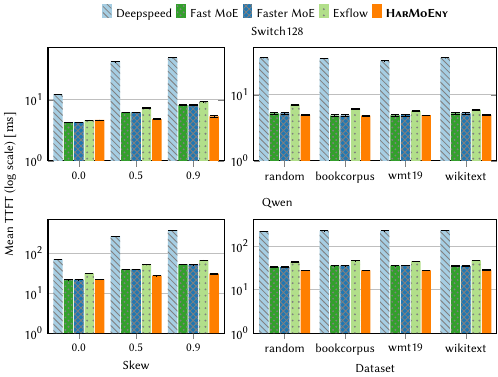}
	\caption{Mean \ac{TTFT} ($\downarrow$ is better) for different systems and different skews (left) and datasets (right) when using the \switch (top) and \qwen (bottom) models.
		}
	\label{fig:skew-mttft}
\end{figure}

\noindent\textbf{Skewed datasets.}
\sys performs especially well when workloads exhibit high expert popularity skew. To showcase this feature, we artificially create workloads with 50\% and 90\% skew, \ie, 50\% and 90\% of the tokens are routed to one popular expert, as detailed above.
\Cref{fig:skew-tput} (left) and \Cref{fig:skew-mttft} (left) show the throughput and mean TTFT for the \textsc{Constant} dataset with different levels of router skew.
In the scenario with 90\% skew, \sys outperforms \fastmoe and \fastermoe by 1.5x and 1.7x for both the \switch and \qwen models, respectively.
\sys beats \deepspeed by 9.1x for the 90\% skew and 8.8x for 50\% skew scenario for \switch.
The trend remains the same for \qwen.
When compared to the \ac{MoE} inference solution \exflow, \sys performs 1.7x and 1.5x better on the 90\% and 50\% skew, respectively for \switch.
The performance boost is even higher for \qwen that has larger experts with \sys outperforming \exflow by 2.2x and 1.8x in the 90\% and 50\% skew scenarios.
In the 50\% skewed workload, \sys is 1.3x and 1.4x faster than \fastermoe for \switch and \qwen, respectively. 

Similar to the baselines, there is a deterioration in the throughput of \sys as the skew increases. 
This happens because in extremely skewed cases, all the experts except for the popular one have very few tokens to process.
Therefore, \sys is not able to efficiently mask the expert fetching with computation leading to slightly lower throughputs.
However, it is important to note that \sys outperforms the state-of-the-art baselines in the skewed scenarios.

As expected, there is little difference between most systems when all experts are equally popular (0\% skew).
\fastmoe and \fastermoe are 8\% faster than \sys in the \switch workload with no skew due to \sys's increased overhead when scheduling experts for each batch.
This overhead is visible as the number of experts in \switch is high (128).
The scheduling compute is not noticeable for \qwen, which uses 60 experts per \ac{MoE} block.
Furthermore, \sys outperforms both \deepspeed and \exflow in terms of throughput and TTFT in the experiments with \qwen without any skew.
The lower performance of \deepspeed is due to the scheduling policy and that of \exflow can be attributed to the inability to adapt the expert placement according to runtime skew.

\vspace{1mm}
\noindent\textbf{Real-world datasets.}
\Cref{fig:skew-tput} (right) and \Cref{fig:skew-mttft} (right) show the throughput and mean TTFT of \sys, \deepspeed, \fastmoe, \fastermoe, and \exflow running the \switch (top) and \qwen (bottom) models.
The figures show the three real-world datasets and the \textsc{random} dataset described in \Cref{sec:datasets}.
Notably, \sys maintains steady high throughput (201 tokens/s and 36 tokens/s for \switch and \qwen respectively) and low mean \ac{TTFT} (5ms and 27ms for \switch and \qwen respectively) in all real-world datasets.
Compared to \exflow and \fastermoe, this results in a speedup of 20\% and 7\% on the \textsc{random} and \textsc{wikitext} datasets, respectively.
\deepspeed remains an outlier here with very low throughput and high TTFT.
This is due to \sys's lightweight load rebalancing mechanism and asynchronous pre-fetching that keep all GPUs operating at close to 100\% (more details in Section~\ref{sec:ablation}).

For the Switch transformer model, \sys is on par with \fastermoe and \fastmoe. Since the size of the experts is relatively small (18MB), \fastmoe's and \fastermoe's expert shadowing mechanism is enough to ensure good load balancing because the GPU memory can comfortably host the shadowed experts. \fastmoe and \fastermoe obtain virtually identical results, within 92\% to 98\% the throughput of \sys across the four workloads. However, the throughput difference is accentuated for larger models. For \qwen (33MB per expert), the gap widens compared to \switch. \sys is 15\% to 28\% faster than \fastmoe and \fastermoe because the extent of the expert shadowing is constrained by the GPU memory. \exflow, while having consistent performance, falls behind \sys for both models because it does not run the expert placement optimization often enough to keep up with expert popularity fluctuations. 

\begin{figure}[t]
    \centering
    \inputplot{plots/peaks}{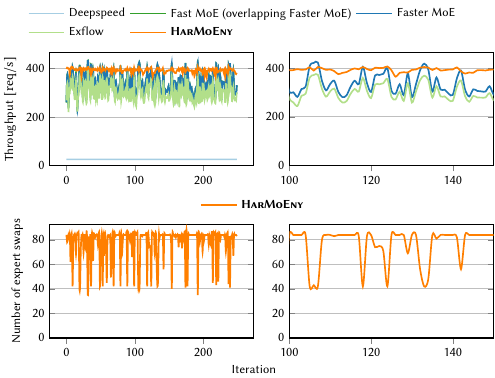}
    \caption{Throughput with 0\%--50\% skew randomly chosen every batch on \switch.  Right-hand zooms into a shorter interval. Swaps in \sys maintain high throughput.}
    \label{fig:systems_throughput_router_dynamic_skew}
\end{figure}

\begin{figure}
    \centering
    \inputplot{plots/timeline_systems}{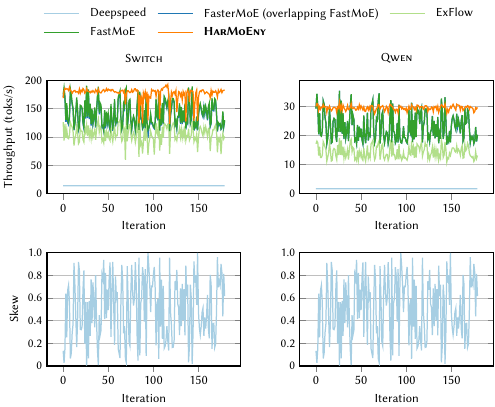}
    \caption{The throughput of \sys and baselines across iterations (top) and skew in expert popularity (bottom), for the \switch (left) and \qwen (right) models. \sys maintains consistent throughput while skew varies from 0\% to 95\% per batch.}
    \label{fig:systems_throughput_router_dynamic_skew_with_qwen}
\end{figure}

\vspace{1mm}
\noindent\textbf{Fluctuating expert popularity over time.} Figure~\ref{fig:systems_throughput_router_dynamic_skew} shows a scenario where each batch has a different randomly chosen expert popularity skew (between 0\% and 50\%).
From the top row, it is evident that \sys maintains high and steady throughput while the other baselines have drops in throughput by up to 37\%.
This scenario shows the effectiveness of \sys's lightweight approach, which can sustain rapidly changing load imbalance without affecting throughput.
\Cref{fig:systems_throughput_router_dynamic_skew} (bottom) shows the number of expert swaps in \sys for every batch of input request.
The dynamic number of expert swaps confirms that the expert pre-fetching mechanism running out of the critical path maintains stable high throughputs.
From the zoomed-in version in \Cref{fig:systems_throughput_router_dynamic_skew} (right column), we can observe that baselines \fastmoe, \fastermoe, and \exflow perform really well and \fastermoe surpasses \sys in batches where \sys swaps the least experts.
These are batches where there is almost negligible load imbalance across the GPUs.
In other words, the baselines work well when there is almost no skew in expert popularity, and hence, all GPUs perform equal computation.
In such scenarios, the lower throughput of \sys is due to the overhead of the token scheduler (see \Cref{sec:scheduler}).

\Cref{fig:systems_throughput_router_dynamic_skew_with_qwen} extends the analysis from \Cref{fig:systems_throughput_router_dynamic_skew} by evaluating a broader range of expert skew values, ranging from 0\% to 95\%.
This figure also includes results for \qwen, and the bottom row annotates the expert skew at each iteration. \sys not only achieves the highest overall throughput but also demonstrates significantly lower variance across batches of the same size but differing skew levels.
Specifically, \sys shows a variance of 152 $toks^2/s^2$ compared to \exflow's 206 $toks^2/s^2$, \fastermoe's 447 $toks^2/s^2$, and \fastmoe's 477 $toks^2/s^2$.
While \deepspeed exhibits the lowest variance, this comes at the cost of throughput due to its input padding strategy, yielding only 13 $tok/s$ versus \sys’s 176 $tok/s$. Therefore, this comparison is not entirely fair. For \qwen, \sys achieves a larger variance reduction, with a variance of just 0.59 $tok^2/s^2$, significantly outperforming \exflow's 4.58 $toks^2/s^2$ (7.76$\times$), \fastermoe's 22.77 $toks^2/s^2$ (38.59$\times$), and \fastmoe's 22.9 $toks^2/s^2$ (38.8$\times$).
In summary, when the expert popularity changes across batches, \sys achieves a high throughput by dynamically adapting to the load imbalance.

\subsection{Ablation study of \sys}

\subsubsection{Time breakdown of \sys components}
\begin{figure}
	\centering
	\inputplot{plots/time_breakdown}{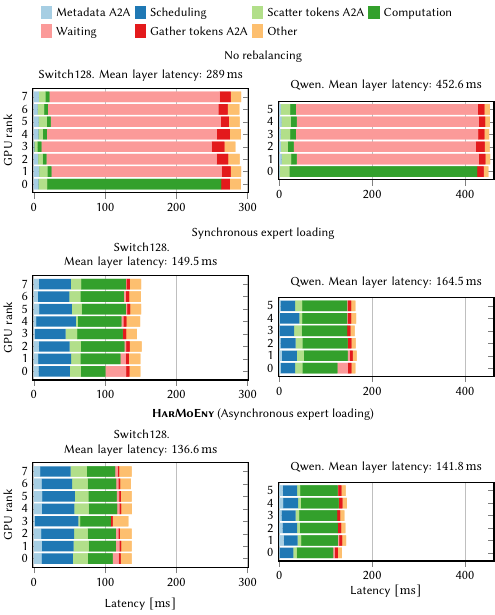}
	\caption{Time breakdown of \sys and baselines for the \switch (left) and \qwen (right) models. }
	\label{fig:time_breakdown}
\end{figure}

We now conduct a time breakdown of the different operations when serving an \ac{MoE} with \sys.
We adopt the \textsc{Constant} workload and assign 90\% tokens to the first 10 experts ($\alpha = 0.9$).
We use NVIDIA CUDA Events to obtain a fine-grained time breakdown of the operations in the first \ac{MoE} layer.
\Cref{fig:time_breakdown} shows the duration of different operations in \sys without any rebalancing (top), with rebalancing but without asynchronous expert loading (middle), and our original \sys with all its components (bottom).
Since the first 10 experts are loaded on GPU 0, we observe significant GPU waiting time for all other GPUs when we do not rebalance the token load using \Cref{alg:scheduling}, which is in line with the discussion in \Cref{sec:load_imbalance}.
Specifically, GPUs 1--7 spend on average 85.7\% (\qwen) of the time waiting for GPU 0 to finish processing experts.

Our token rebalancing algorithm shrinks the waiting time on all GPUs.
The mean waiting time goes from 82.6\% and 85.7\% of the total GPU time down to a mere 2.6\% and 1\% for \switch and \qwen, respectively across all GPUs, as seen in \Cref{fig:time_breakdown} (middle).
For the \switch model, token rebalancing reduces the mean layer latency from \SI{289}{\milli\second} to \SI{149.5}{\milli\second}, a total reduction of 48.3\%.
This reduction is even more pronounced with the \qwen model: 63.7\% compared to when not rebalancing tokens.
On average, our scheduling algorithm takes, 30.8\% and 20.3\% of the mean latency for the \switch and \qwen model, respectively.
While scheduling and rebalancing in \sys takes time, it brings a significant decrease in total latency.

\Cref{fig:time_breakdown} (bottom) shows the timeline of operations of \sys with all components enabled.
Asynchronous expert loading further reduces the latency to \SI{136.6}{\milli\second} (-8.63\% over synchronous loading)  for \switch and to \SI{141.8}{\milli\second} (-13.8\% over synchronous loading) for \qwen.
Thus, we conclude that the combination of token rescheduling and asynchronous fetching in \sys effectively minimizes the idling of GPUs and reduces the latency of \ac{MoE} inference.

\subsubsection{Load balancing policies}
\label{sec:ablation}

\begin{figure}[t]
	\centering
	\inputplot{plots/skewnew}{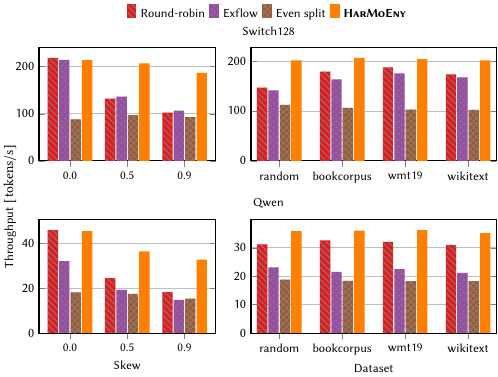}
	\caption{Throughput ($\uparrow$ is better) for different token load policies and different skews (left) and datasets (right) when using the \switch (top) and \qwen models (bottom). 
		}
	\label{fig:skewnew-tput}
\end{figure}

We next experiment with \sys and when using different token rebalancing policies and measure the throughput in tokens/s. We implement the token rebalancing policies on top of \sys to account for differences in system baselines implementations.

\label{sec:eval_policies}
We evaluate \sys against three other policies for token routing: 
\begin{enumerate*}[label=(\arabic*)]
    \item With the Round-robin policy, tokens are sent to the GPU housing the specified expert, regardless of imbalance, with experts distributed to GPUs in a round-robin manner.
    This policy is employed by \deepspeed, \fastmoe, and \fastermoe;
    \item The \exflow policy utilizes an integer programming-based approach to optimize token scheduling and routing by modeling the inter-layer affinity between tokens and experts, formulating a placement optimization problem to minimize the communication cost of routing tokens between GPUs;
    \item Even Split evenly distributes the tokens for each expert to each GPU and replicates all experts on all GPUs.
For example, if there are $a$ tokens for expert 0 and four GPUs then $\frac{a}{4}$ tokens will be sent to each of the four GPUs.
This achieves a perfect load balance across all experts at the cost of replication of all experts across all the GPUs. 
\end{enumerate*}

\begin{figure}[t]
	\centering
	\inputplot{plots/skewnew-mttft}{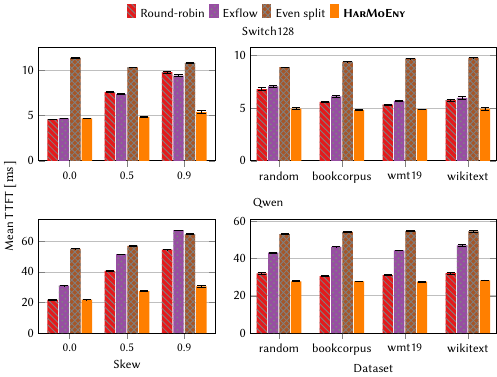}
	\caption{Mean \ac{TTFT} ($\downarrow$ is better) for different token load policies and different skews (left) and datasets (right) when using the \switch (top) and \qwen models (bottom).
		}
	\label{fig:skewnew-mttft}
\end{figure}

\vspace{1mm}
\noindent\textbf{Skewed datasets.}
\Cref{fig:skewnew-tput} (left) and \Cref{fig:skewnew-mttft} (left) show the throughput and mean \ac{TTFT} under different token load policies and different skew levels $ \alpha $, for the two \ac{MoE} models.
In these experiments, we skew the load of a single expert.
\sys, when using \switch with $\alpha = 0$ (no skew), reaches a throughput of 213 tokens/s which is comparable to that of the Round-robin and \exflow policies.
However, as $ \alpha $ increases and consequentially, the load imbalance, \sys reaches a significantly higher throughput than the other policies.
For $ \alpha = 0.9 $, \sys reaches a throughput of 186 tokens/s compared to 106 tokens/s for \exflow, the best-performing baseline.
We observe similar trends when using the \qwen model (\Cref{fig:skewnew-tput}, bottom left).
Even though the even split policy achieves perfect load balance, its performance is relatively low.
This is because each GPU has to load and execute every expert, thus increasing the time taken to process a batch of inference requests.

\Cref{fig:skewnew-mttft} (left) shows that for $ \alpha = 0 $ and with the \switch model, \sys has a mean \ac{TTFT} of \SI{4.68}{\milli\second}, which is comparable to that of the Round-robin and \exflow policies.
However, when $\alpha $ increases, so does the mean \ac{TTFT} of other policies.
With $ \alpha = 0.9 $ and with the \switch model, \sys has a mean \ac{TTFT} of \SI{5.36}{\milli\second}, compared to \SI{9.77}{\milli\second} and \SI{9.38}{\milli\second} for the Round-robin and \exflow policies, respectively.
The mean \ac{TTFT} of \sys is also competitive when using the \qwen model.
Thus, \sys exhibits excellent mean \ac{TTFT}, even under heavy token imbalances.

\noindent\textbf{Real-world datasets.}
\Cref{fig:skewnew-tput} (right) and \Cref{fig:skewnew-mttft} (right) show the throughput and mean \ac{TTFT} of \sys and other policies for real-world datasets.
For all datasets and models used, \sys reaches the highest throughput.
This is the most pronounced when using the \textsc{random} dataset and \switch model.
\Cref{fig:skewnew-mttft} (right) shows that 
\sys exhibits the lowest mean \ac{TTFT}, thus justifying the token distribution policy in \sys.

\section{Related work}
\label{sec:related}

In this section, we discuss the related work and present the comparison with \sys.

\paragraph{\textbf{Efficient MoE inference.}}
Recent works address the efficiency problem for MoE inference by optimizing communication overhead, token imbalance, and GPU kernels.
DeepSpeed-MoE Inference~\cite{rajbhandari2022deepspeed} is a framework for serving MoE models providing flexible parallelization combinations, highly optimized MoE-related kernels, and an efficient communication subsystem.
However, the static expert assignment in DeepSpeed-MoE struggles under skewed and dynamic workloads.
Tutel improves upon DeepSpeed-MoE by introducing adaptive parallelism at runtime~\cite{hwang2023tutel}.
While Tutel handles dynamic workloads better than DeepSpeed-MoE, its static expert placement limits its ability to handle severe load imbalance.
ExFlow reduces the all-to-all communication overhead by exploiting inter-layer expert affinity~\cite{yao2024exploiting}.
While effective, it assumes stable expert routing patterns, which may not adapt well to fluctuating inputs. 
Lina~\cite{li2023accelerating} improves inference performance through dynamic resource scheduling to balance skewed workloads with a 2-phase scheme by profiling experts and predicting expert selection.
Though effective in scenarios where inference requests are from similar domains, Lina needs to reallocate resources when the expert popularity changes.
In contrast, \sys directly targets load imbalance in skewed workloads through token redistribution and dynamic expert placement.
In addition, frameworks like DeepSpeed-MII~\cite{deepspeed-mii} and vLLM~\cite{kwon2023efficient} are under active development and utilize highly optimized CUDA kernels targeted to specific GPU architecture.
Our approach is orthogonal to these.
\sys works at the application layer and can be further optimized with specialized GPU kernels as in DeepSpeed-MII or vLLM.

\paragraph{\textbf{Efficient MoE training.}}
Several frameworks have been proposed to make training of MoE models efficient.
FastMoE~\cite{he2021fastmoe} introduces an MoE training system with hierarchical interfaces and optimized CUDA kernels, enabling scalability but relying on static expert placement.
FasterMoE~\cite{he2022fastermoe} builds on this by addressing load imbalance through dynamic shadowing and fine-grained scheduling, while introducing congestion-avoiding expert selection during training.
Since the experts are sparsely activated, Megablocks~\cite{gale2023megablocks} achieves hardware efficiency by combining expert computations into block-sparse operations.
Similar to DeepSpeed-MoE, SmartMoE~\cite{zhai2023smartmoe} supports dynamic and hybrid parallelization strategies for MoE training.
Finally, Prophet~\cite{wang2023prophet} utilizes a fine-grained planner and exploits similarity in token distribution across training iterations.
Contrary to the aforementioned approaches, \sys focuses on dynamic inference load imbalance rather than training.

\section{Conclusion}
\label{sec:conclusion}

We presented \sys, a novel system that addresses load imbalance in multi-GPU inference of \ac{MoE} models.
Through the combination of dynamic token rebalancing and asynchronous expert fetching, \sys achieves near-perfect load balancing, significantly reducing inference latency.
Our comprehensive evaluation using multiple datasets and \ac{MoE} models demonstrated that \sys outperforms state-of-the-art baselines.
With heavy token imbalance, \sys increases throughput by up to 70.1\% and reduces time-to-first-token by up to 41.1\%, compared to the next-best competitor, while maintaining stable throughput.

\appendix

\section{Appendix: Model Statistics}
\label{app:model_specs}

\begin{table}[ht]
    \centering
    \begin{tabular}{c|c|c|c}
        \toprule
        Model & MoE Layers & Experts & Expert size (MB) \\
        \midrule
        \switch & 12 & 128 & 18 \\
        \qwen & 24 & 60 & 33 \\
        \bottomrule
    \end{tabular}
    \caption{Specifications of the \ac{MoE} models used in  evaluation.}
    \label{tab:moe-configs}
\end{table}

\Cref{tab:moe-configs} shows the specifications of the two \ac{MoE} models we have used in our evaluation.

\section{Appendix: Mathematical Estimate for $q$}
\label{app:q_inequality}
Following is a step-by-step breakdown of getting a formula for estimating the minimal size for $q$ given that the expert is a 2-layer MLP.
$q$ represents the number of tokens that are required so that its processing time is greater than the time it takes to load an expert. Let the expert have two linear layers with the first being of size $m \times p$, and the second being $p \times m$. The expert is evaluated as $xW^1W^2$. This can be represented as:

\begin{align}
    \frac{\text{Number of Floating Point Operations}}{\text{GPU FLOPS}} &> \frac{\text{Expert Size}}{\text{PCIe Bandwidth}} \\
    \frac{|O|}{\phi} &> \frac{|E|}{\beta} \\
    \frac{qp(2m-1)+qm(2p-1)}{\phi} &> \frac{(mp+pm)d_{type}}{\beta} \\
\end{align}
On ignoring small terms and simplifying, we get:
\begin{align}
    \frac{2qpm+2qpm}{\phi} &> \frac{2pm \cdot d_{type}}{\beta} \\
    \frac{q\cdot4pm}{\phi} &> \frac{2pm \cdot d_{type}}{\beta} \\
    q &> \frac{\phi \cdot d_{type}}{2\beta}
\end{align}

\section{Integrating \sys}
\label{app:integration}
The following Python code demonstrates how to add \sys to an existing model. The \texttt{replace\_moe\_layer} function injects our MoE implementation based on user-specified parameters.
\begin{lstlisting}[language=Python]
	from harmonymoe.utils import replace_moe_layer
	from harmonymoe.moe_layer import MoEConfig, MoELayer
	
	model = create_pytorch_model() # Custom model
	
	config = MoEConfig(
        rank,
        world_size,
        scheduling_policy,
        expert_cache_size,
        eq_tokens,
        d_model,
        num_experts,
	)
	
	replace_moe_layer(
    	model,
    	moe_parent_type, 
    	moe_type, 
    	path_to_experts, 
    	path_to_router_linear_layer,
    	config,
	)
\end{lstlisting}

\bibliographystyle{plain}
\bibliography{main.bib}

\begin{thebibliography}{10}

\bibitem{achiam2023gpt}
Josh Achiam, Steven Adler, Sandhini Agarwal, Lama Ahmad, Ilge Akkaya,
  Florencia~Leoni Aleman, Diogo Almeida, Janko Altenschmidt, Sam Altman,
  Shyamal Anadkat, et~al.
\newblock {GPT}-4 technical report.
\newblock {\em arXiv:2303.08774}, 2023.

\bibitem{aws-inference}
Jeff Barr.
\newblock Amazon {EC2} update – inf1 instances with {AWS} inferentia chips
  for high performance cost-effective inferencing, 2019.
\newblock Accessed: January 2025.

\bibitem{bengio2000neuralprobabilisticlanguagemodel}
Yoshua Bengio, R\'{e}jean Ducharme, and Pascal Vincent.
\newblock A neural probabilistic language model.
\newblock In {\em NeurIPS}, 2000.

\bibitem{bianchini2024datacenter}
Ricardo Bianchini, Christian Belady, and Anand Sivasubramaniam.
\newblock Datacenter power and energy management: past, present, and future.
\newblock {\em IEEE Micro}, 2024.

\bibitem{brown2020language}
Tom Brown, Benjamin Mann, Nick Ryder, Melanie Subbiah, Jared~D Kaplan, Prafulla
  Dhariwal, Arvind Neelakantan, Pranav Shyam, Girish Sastry, Amanda Askell,
  et~al.
\newblock Language models are few-shot learners.
\newblock {\em NeurIPS}, 2020.

\bibitem{dean2012}
Jeffrey Dean, Greg Corrado, Rajat Monga, Kai Chen, Matthieu Devin, Mark Mao,
  Marc\textquotesingle~aurelio Ranzato, Andrew Senior, Paul Tucker, Ke~Yang,
  Quoc Le, and Andrew Ng.
\newblock Large scale distributed deep networks.
\newblock In {\em NeurIPS}, volume~25, 2012.

\bibitem{delimitrou2014quasar}
Christina Delimitrou and Christos Kozyrakis.
\newblock Quasar: Resource-efficient and qos-aware cluster management.
\newblock {\em ACM Sigplan Notices}, 49(4), 2014.

\bibitem{fedus2022switchtransformers}
William Fedus, Barret Zoph, and Noam Shazeer.
\newblock Switch transformers: Scaling to trillion parameter models with simple
  and efficient sparsity.
\newblock {\em Journal of Machine Learning Research}, 23(120), 2022.

\bibitem{wikimedia2019wmt19}
Wikimedia Foundation.
\newblock Acl 2019 fourth conference on machine translation (wmt19), shared
  task: Machine translation of news.

\bibitem{gale2023megablocks}
Trevor Gale, Deepak Narayanan, Cliff Young, and Matei Zaharia.
\newblock Megablocks: Efficient sparse training with mixture-of-experts.
\newblock In {\em MLSys}, 2023.

\bibitem{gupta2019characterbasednmttransformer}
Rohit Gupta, Laurent Besacier, Marc Dymetman, and Matthias Gallé.
\newblock Character-based {NMT} with transformer.
\newblock {\em arXiv:1911.04997}, 2019.

\bibitem{he2021fastmoe}
Jiaao He, Jiezhong Qiu, Aohan Zeng, Zhilin Yang, Jidong Zhai, and Jie Tang.
\newblock Fastmoe: A fast mixture-of-expert training system.
\newblock {\em arXiv:2103.13262}, 2021.

\bibitem{he2022fastermoe}
Jiaao He, Jidong Zhai, Tiago Antunes, Haojie Wang, Fuwen Luo, Shangfeng Shi,
  and Qin Li.
\newblock Fastermoe: Modeling and optimizing training of large-scale dynamic
  pre-trained models.
\newblock In {\em PPoPP}, 2022.

\bibitem{hwang2023tutel}
Changho Hwang, Wei Cui, Yifan Xiong, Ziyue Yang, Ze~Liu, Han Hu, Zilong Wang,
  Rafael Salas, Jithin Jose, Prabhat Ram, et~al.
\newblock Tutel: Adaptive mixture-of-experts at scale.
\newblock {\em MLSys}, 2023.

\bibitem{jacobs1991adaptive}
Robert~A Jacobs, Michael~I Jordan, Steven~J Nowlan, and Geoffrey~E Hinton.
\newblock Adaptive mixtures of local experts.
\newblock {\em Neural computation}, 3(1), 1991.

\bibitem{jiang2024mixtralexperts}
Albert~Q. Jiang, Alexandre Sablayrolles, Antoine Roux, Arthur Mensch, Blanche
  Savary, Chris Bamford, Devendra~Singh Chaplot, Diego de~las Casas, Emma~Bou
  Hanna, Florian Bressand, Gianna Lengyel, Guillaume Bour, Guillaume Lample,
  Lélio~Renard Lavaud, Lucile Saulnier, Marie-Anne Lachaux, Pierre Stock,
  Sandeep Subramanian, Sophia Yang, Szymon Antoniak, Teven~Le Scao, Théophile
  Gervet, Thibaut Lavril, Thomas Wang, Timothée Lacroix, and William~El Sayed.
\newblock Mixtral of experts.
\newblock {\em arXiv:2401.04088}, 2024.

\bibitem{kaplan2020scalinglawsneurallanguage}
Jared Kaplan, Sam McCandlish, Tom Henighan, Tom~B. Brown, Benjamin Chess, Rewon
  Child, Scott Gray, Alec Radford, Jeffrey Wu, and Dario Amodei.
\newblock Scaling laws for neural language models.
\newblock {\em arXiv:2001.08361}, 2020.

\bibitem{krizhevsky2009learning}
Alex Krizhevsky.
\newblock Learning multiple layers of features from tiny images.
\newblock Technical report, University of Toronto, 2009.

\bibitem{kudo2018subwordregularizationimprovingneural}
Taku Kudo.
\newblock Subword regularization: Improving neural network translation models
  with multiple subword candidates.
\newblock In {\em ACL}, 2018.

\bibitem{kwon2023efficient}
Woosuk Kwon, Zhuohan Li, Siyuan Zhuang, Ying Sheng, Lianmin Zheng, Cody~Hao Yu,
  Joseph~E. Gonzalez, Hao Zhang, and Ion Stoica.
\newblock Efficient memory management for large language model serving with
  {PagedAttention}.
\newblock In {\em SOSP}, 2023.

\bibitem{nvidia-inference}
George Leopold.
\newblock {AWS} to offer nvidia’s t4 {GPUs} for {AI} inferencing, 2019.
\newblock Accessed: January 2025.

\bibitem{lepikhin2021gshard}
Dmitry Lepikhin, HyoukJoong Lee, Yuanzhong Xu, Dehao Chen, Orhan Firat, Yanping
  Huang, Maxim Krikun, Noam Shazeer, and Zhifeng Chen.
\newblock Gshard: Scaling giant models with conditional computation and
  automatic sharding.
\newblock In {\em ICLR}, 2021.

\bibitem{li2023accelerating}
Jiamin Li, Yimin Jiang, Yibo Zhu, Cong Wang, and Hong Xu.
\newblock Accelerating distributed {MoE} training and inference with lina.
\newblock In {\em USENIX ATC}, 2023.

\bibitem{liu2024deepseek}
Aixin Liu, Bei Feng, Bing Xue, Bingxuan Wang, Bochao Wu, Chengda Lu, Chenggang
  Zhao, Chengqi Deng, Chenyu Zhang, Chong Ruan, et~al.
\newblock Deepseek-v3 technical report.
\newblock {\em arXiv:2412.19437}, 2024.

\bibitem{merity2016pointer}
Stephen Merity, Caiming Xiong, James Bradbury, and Richard Socher.
\newblock Pointer sentinel mixture models, 2016.

\bibitem{deepspeed-mii}
Microsoft.
\newblock Deepspeed-mii: Mii makes low-latency and high-throughput inference
  possible, powered by deepspeed.
\newblock \url{https://github.com/microsoft/DeepSpeed-MII}, 2022.
\newblock Accessed: 2025-01-13.

\bibitem{torchSoftware}
Adam Paszke, Sam Gross, Francisco Massa, Adam Lerer, James Bradbury, Gregory
  Chanan, Trevor Killeen, Zeming Lin, Natalia Gimelshein, Luca Antiga, et~al.
\newblock Pytorch: An imperative style, high-performance deep learning library.
\newblock In {\em NeurIPS}, 2019.

\bibitem{raffel2020exploring}
Colin Raffel, Noam Shazeer, Adam Roberts, Katherine Lee, Sharan Narang, Michael
  Matena, Yanqi Zhou, Wei Li, and Peter~J Liu.
\newblock Exploring the limits of transfer learning with a unified text-to-text
  transformer.
\newblock {\em Journal of machine learning research}, 21(140), 2020.

\bibitem{rajbhandari2022deepspeed}
Samyam Rajbhandari, Conglong Li, Zhewei Yao, Minjia Zhang, Reza~Yazdani
  Aminabadi, Ammar~Ahmad Awan, Jeff Rasley, and Yuxiong He.
\newblock {DeepSpeed}-{MoE}: Advancing mixture-of-experts inference and
  training to power next-generation ai scale.
\newblock In {\em ICML}, 2022.

\bibitem{shazeer2017outrageouslylargeneuralnetworks}
Noam Shazeer, Azalia Mirhoseini, Krzysztof Maziarz, Andy Davis, Quoc Le,
  Geoffrey Hinton, and Jeff Dean.
\newblock Outrageously large neural networks: The sparsely-gated
  mixture-of-experts layer.
\newblock In {\em ICLR}, 2017.

\bibitem{shoeybi2020megatronlmtrainingmultibillionparameter}
Mohammad Shoeybi, Mostofa Patwary, Raul Puri, Patrick LeGresley, Jared Casper,
  and Bryan Catanzaro.
\newblock Megatron-{LM}: Training multi-billion parameter language models using
  model parallelism, 2020.

\bibitem{tirmazi2020borg}
Muhammad Tirmazi, Adam Barker, Nan Deng, Md~E Haque, Zhijing~Gene Qin, Steven
  Hand, Mor Harchol-Balter, and John Wilkes.
\newblock Borg: the next generation.
\newblock In {\em EuroSys}, 2020.

\bibitem{vaswani2017attention}
A~Vaswani et~al.
\newblock Attention is all you need.
\newblock {\em NeurIPS}, 2017.

\bibitem{wang2023prophet}
Wei Wang, Zhiquan Lai, Shengwei Li, Weijie Liu, Keshi Ge, Yujie Liu, Ao~Shen,
  and Dongsheng Li.
\newblock Prophet: Fine-grained load balancing for parallel training of
  large-scale moe models.
\newblock In {\em IEEE International Conference on Cluster Computing
  (CLUSTER)}, 2023.

\bibitem{xu2024resource}
Mengwei Xu, Dongqi Cai, Wangsong Yin, Shangguang Wang, Xin Jin, and Xuanzhe
  Liu.
\newblock Resource-efficient algorithms and systems of foundation models: A
  survey.
\newblock {\em ACM Computing Surveys}, 2024.

\bibitem{yang2024qwen2}
An~Yang, Baosong Yang, Beichen Zhang, Binyuan Hui, Bo~Zheng, Bowen Yu,
  Chengyuan Li, Dayiheng Liu, Fei Huang, Haoran Wei, et~al.
\newblock Qwen2. 5 technical report.
\newblock {\em arXiv:2412.15115}, 2024.

\bibitem{yang2024harnessing}
Jingfeng Yang, Hongye Jin, Ruixiang Tang, Xiaotian Han, Qizhang Feng, Haoming
  Jiang, Shaochen Zhong, Bing Yin, and Xia Hu.
\newblock Harnessing the power of {LLMs} in practice: A survey on chatgpt and
  beyond.
\newblock {\em ACM Transactions on Knowledge Discovery from Data}, 18(6), 2024.

\bibitem{yao2024exploiting}
Jinghan Yao, Quentin Anthony, Aamir Shafi, Hari Subramoni, and Dhabaleswar K~DK
  Panda.
\newblock Exploiting inter-layer expert affinity for accelerating
  mixture-of-experts model inference.
\newblock In {\em IEEE IPDPS}, 2024.

\bibitem{zhai2023smartmoe}
Mingshu Zhai, Jiaao He, Zixuan Ma, Zan Zong, Runqing Zhang, and Jidong Zhai.
\newblock {SmartMoE}: Efficiently training {Sparsely-Activated} models through
  combining offline and online parallelization.
\newblock In {\em USENIX ATC}, 2023.

\bibitem{zhu2015moviebook}
Yukun Zhu, Ryan Kiros, Richard Zemel, Ruslan Salakhutdinov, Raquel Urtasun,
  Antonio Torralba, and Sanja Fidler.
\newblock Aligning books and movies: Towards story-like visual explanations by
  watching movies and reading books.
\newblock In {\em arXiv:1506.06724}, 2015.

\end{thebibliography}

\end{document}